\documentstyle[11pt,psfig,aas2pp4]{article}
\begin{document}


\title{\bf RADIO-QUIET RED QUASARS}

\bigskip
\author{Dong-Woo Kim$^1$}
\affil{Chungnam National University, Taejon, 305-764,
South Korea}
\altaffiltext{1}{also at Harvard-Smithsonian Center for
Astrophysics, 60 Garden Street, Cambridge, MA 02138}

\author{Martin Elvis}
\affil{Harvard-Smithsonian Center for Astrophysics, 60
Garden Street, Cambridge, MA 02138}

\centerline{\tt version: 1pm, November 9 1998}

\begin{abstract}

We have performed a successful targetted search for a population
of red radio-quiet, and probably absorbed, quasars. Radio-quiet,
optically-red ROSAT PSPC X-ray sources brighter than $1 \times
10^{-13} $ erg~cm$^{-2}$~s$^{-1}$ were searched for red
(O-E$>$2.0, O $\leq$20) counterparts in the APM catalog of
Palomar Sky Survey objects. Of 45 objects for which we obtained
adequate follow-up optical spectroscopy, we have found 7 red
quasars, 5 with $\alpha_{opt}<-2$. Their redshifts range from
0.06 to 0.31, and their luminosities are moderate, lying on the
Seyfert/Quasar boundary. These red quasars strengthen the case
for a radio-quiet population that is the counterpart of the
radio-loud red quasars found by Smith and Spinrad (1980), and
Webster et al. (1995). Unidentified, fainter, sources could
increase the fraction of red quasars by up to a factor 7.

For the red quasars found here, the $H\alpha$/$H\beta$ ratios,
optical slope and X-ray colors all indicate that they are
absorbed by $A_V\sim 2$, rather than having intrinsically red
spectra. This amount of obscuration seems to hide $\sim$1-7\% of
quasars at a given observed flux, or $\sim$3-20\% when their
fluxes are corrected to their intrinsic values. This size of
population is consistent with earlier limits, with predicted
values from Comastri et al. (1995), and is comparable to the rate
found among radio-loud quasars.

A large population of more heavily absorbed ($A_V=5$), fainter,
quasars equal in size to the blue population could exist, without
violating existing upper limits, in accord with the Comastri et
al. (1995) predictions.

\end{abstract}

\newpage
\centerline {\bf 1. Introduction}
\medskip

Quasars are the canonical ``ultraviolet excess'' objects (Sandage
1965). Yet red quasars have been found in radio-selected samples
by Smith and Spinrad (1980) and Webster et al. (1995).  Webster
et al. (1995) proposed that a large fraction, perhaps 80\%, of
radio-loud quasars might have been hidden as red objects.
Moreover, the currently favored explanations for the cosmic X-ray
background invoke a population of heavily obscured active
galactic nuclei (AGNs) 5 times more common than the unobscured
population (Comastri et al., 1995).  If the small number of known
red quasars really are the `tip of the iceberg' of a large, even
dominant, quasar population then the consequences would be
interesting: the overall AGN population - and so the massive
black hole population - may be five times larger than had been
thought; obscured quasars would be a long-lived evolutionary
phase (c.f. Sanders et al., 1988), or all quasars may be hidden
along 80\% of possible lines of sight; and red quasars may
contribute importantly to the cosmic X-ray background.  The
Webster et al. conclusion is widely disputed.  Boyle \& diMatteo
(1995), Stickel et al. (1996) and Benn et al., (1998) all argue
that any missing population must be smaller, and perhaps
insignificant, while Gunn \& Shanks (1998) disagree.  Here we
present an X-ray based survey targetted explicity at red AGN, to
find radio-quiet red quasars.

Radio-loud red quasars are relatively easy to find, since the
radio emission is unaffected by absorbing gas or dust and, in low
frequency surveys, usually comes from the large radio lobes that
lie well outside any obscuring material in the host galaxy. An
explicit, albeit small-scale, search for radio- and X-ray-loud
but optically-quiet quasars, which should include reddened
quasars was, however not successful (Kollgaard et al., 1995).

Radio-quiet red quasars are much harder to find, although they
might be expected to be much more common. In the normal
unreddened population radio-quiet quasars outnumber radio-loud
quasars 10 to one (e.g. `EMSS' Stocke et al. 1991; `PG' Kellerman
et al., 1989).  However, most optical quasar surveys are actively
biased against finding red quasars. Since these surveys primarily
search for UV bright objects (e.g. `Markarian', Lipovetsky,
Markarian \& Stepanian 1987; `PG', Schmidt and Green 1981; `LBQS'
, Hewett et al. 1995; `HS' Engels et al., 1998) they are blind to
red objects. As a result optical bounds on how large a population
of radio-quiet red quasars might exist are weak.

X-ray selection provides a way of selecting red quasars
efficiently: hard X-rays (2-10~keV) penetrate even 10s of
magnitudes of optical extinction with minimal absorption. Even
the lower energy band of ROSAT (0.5-2.5~keV) is not strongly
affected by optical extinction of up to $\sim$2~magnitudes
\footnote{For standard Milky Way composition and dust-to-gas
ratio (Bohlin, Savage \& Drake 1978, Seaton 1979) the PSPC count
rate is reduced by a factor $1/e$ at $A_V$=1.7 ($N_H=3\times
10^{21}$atoms~cm$^{-2}$), for a power-law photon index of 2.0,
all at zero redshift.}
. Astrophysics might be against us since, although blue quasars
are overwhelmingly X-ray loud (Avni and Tananbaum 1986), red ones
might be intrinsically X-ray faint. Fortunately though, radio-loud
red quasars are known to be X-ray sources (Bregman et al., 1985,
Elvis et al., 1994), so this is unlikely to be a problem.

Complete flux limited X-ray surveys have found some red quasars
or AGN (Stocke et al., 1982, Kruper \& Canizares 1989,
Puchnarewicz et al., 1996, 1997). In particular the RIXOS survey
(Puchnarewicz \& Mason 1998) has identified a small sample of red
quasars.

As part of a more general search for minority populations of
X-ray sources we have used the ROSAT (Tr\"{u}mper et al. 1983)
archive of pointed PSPC (Pfefferman et al., 1987) data to design
an efficient search strategy explicitly targetting red quasars.
The ROSAT pointed archive provides us with a tenfold increase in
the number of X-ray sources from before ROSAT (to about 70,000),
and reaches more than ten times fainter than the ROSAT All Sky
Survey (Voges et al. 1996).  By carefully selecting a small
number of interesting objects we can make an efficient search for
a radio-quiet population of red quasars.

In this paper we find a substantial population of radio-quiet red
AGN on the quasar/Seyfert luminosity boundary.

\bigskip
\centerline {\bf 2. Sample Selection}
\medskip

Most ROSAT sources at high Galactic latitude are blue, unobscured
AGN (e.g. Boyle, Wilkes \& Elvis, 1997, Schmidt et al., 1997). We
use this fact to efficiently select {\em against} such objects
and so isolate any red AGN that may be in the ROSAT archive.

As a compendium of the X-ray sources found by the ROSAT PSPC we
have used the `WGACAT' catalog (which is named after its authors,
White, Giommi \& Angelini 1995).  This catalog was generated from
ROSAT PSPC pointed observations, using a sliding cell detect
algorithm. This method is sensitive to finding point sources, but
can also find spurious sources where extended emission is
present. WGACAT includes a quality flag that notes such dubious
detections based on a visual inspection of the fields. We have
only used X-ray sources with high detection quality to exclude
the spurious sources. From the WGACAT catalog, we have selected
sources by the following criteria:
\begin{enumerate}
\item X-ray bright ($f_X > 10^{-13}$erg~cm$^{-2}$s$^{-1}$), to
allow follow-up observations with other X-ray telescopes;

\item well-detected with signal-to-noise ratio $>$10 and WGACAT
quality flag, DQFLAG $>$5;

\item within r=18~arcmin from the detector center, to provide good
positions; 

\item at high galactic latitude ($|b|>20^{\circ}$), to minimise
the fraction of Galactic stars (which are also red);

\item not within 2~arcmin of the target position (at which point
the source density reaches the background level), to select only
random, serendipitous, sources; and

\item North of decl.=$-18^{\circ}$ in order to have two band
measurements in the Automated Plate-measuring Machine (APM)
catalog (McMahon \& Irwin, 1992), hence giving an archival
optical color.

\item unidentified, with WGACAT class=9999 and no SIMBAD or NED
identification
\footnote{Although only `unidentified' sources were selected, one
(1WGA~J1118.0+4505, see Table 1) turned out to be a known
Seyfert~1 (Bade et al. 1995).}
;

\end{enumerate}

The first six criteria selected 1624 sources.  Since these
sources were selected purely on their X-ray properties they form
a well-defined sample from which to study the incidence of
minority X-ray populations, including any radio-quiet red
quasars. Adding the requirement that a source be unidentified
left 940 X-ray sources which could be examined for having red
optical counterparts. 

We then searched the APM catalog of objects detected on the
Palomar Sky Survey for optical counterparts to the unidentified
X-ray sources.  To find counterparts we used a search radius of
26~arcsec, which corresponds to about 95\% confidence for X-ray
sources within 18~arcmin of the PSPC detector center (Boyle et
al., 1995a). 881 sources had APM catalog counterparts brighter
than the limiting magnitudes O$<$21.5 and E$<$20. (The remaining
`blank' fields are the subject of another study, Elvis, Kim \&
Nicastro 1999, in preparation.)  Among these 881 many had only
O-band magnitudes (suggesting that they are blue), leaving 575
with O-E colors available.

The combination of the ROSAT X-ray flux and the APM magnitudes
allows us to create a rough classification of the X-ray sources
in our sample.  The two Palomar O (blue) and E (red) magnitudes
are close to Johnson B and Cousins R, respectively (Gregg et
al. 1996).  From the X-ray flux and the two optical magnitudes we
can construct a two-color diagram of $\alpha_{ox} -
(O-E)$. $\alpha_{ox}$ is defined by the power law index between 2
keV and 2500\AA\ (Tananbaum et al. 1979; Stocke et
al. 1991). This diagram allows us to select red objects and then
to reduce the Galactic stellar population among these red objects
by selecting the X-ray loud population. Here we make use of the
observation that in stellar sources the X-ray flux for a given
optical flux is much weaker than in AGNs and clusters (e.g.,
Maccacaro et al. 1988).  Figure 1a shows Einstein Medium
Sensitivity Survey (EMSS) X-ray sources (Stocke et al. 1991) in
the $\alpha_{ox}~-~(O-E)$ plane.  This plot clearly illustrates
the distinction between Galactic stellar sources and
extragalactic sources.

Based on the EMSS source distribution, we divided the ROSAT-APM
sources on the same plane (Figure 1b). The O-E APM colors are not
as accurate as the EMSS values, which are based on CCD
photometry. As a result the spread of observed colors is wider
(Figure 1b), and there will be some blue objects in the red zone
and vice versa. To create our list of red quasar candidates we
first excluded the 128 sources with $\alpha_{ox} > 1.8$, because
they are likely to be Galactic stars. Then we excluded another
360 sources with blue colors, $O-E < 2$, because they are most
likely just normal blue unobscured AGN.  This results in a final
sample of 87 X-ray sources defined by the lower-right corner of
figure~1b in the $\alpha_{ox}~-~(O-E)$ plane.

Our sample of 87 optically red X-ray-loud sources is a mere 0.1\%
of the $\sim$70,000 WGACAT sources. The fraction of X-ray sources
that may be red quasars though is much larger: $\sim$5\% of the
initial X-ray selected sample; $\sim$15\% of the unidentified
sources with APM colors; and $\sim$20\% of X-ray bright objects
with APM colors which will, primarily, be AGN.

However, other classes of X-ray source than red quasars can
inhabit this region of the $\alpha_{ox}$~-~$(O-E)$ plane, for
example first ranked elliptical galaxies in distant clusters of
galaxies.  Optical spectroscopy is needed to find red quasars.
We have taken spectra for 51 of the 87 red quasar candidates, as
described in the next section.

\bigskip
\centerline {\bf 3. Observations}
\medskip

\centerline {\em 3.1 Optical Spectroscopy}
\medskip

A typical X-ray error circle containes just 1-2 optical objects
in the APM catalog. Since we have selected against blue
counterparts, we began by observing the brightest red
counterpart. If two objects were present we aligned the
spectrograph slit to obtain spectra of both at once. If this the
first spectra did not find an AGN we then observed the next
faintest, if present.  Since the density of (blue) AGN at B=21 is
only $\sim$0.005 per error circle (e.g. Zitelli et al., 1992) we
expect only 1/4 chance AGN coincidences in the 51 spectra, so
stopping once an AGN is found will not produce a significant
number of false identifications.

We performed optical spectroscopy with MMT on 1997 March 13-15,
with the FLWO 60$^{\prime\prime}$ telescope on 1996 November
16-17 and 1997 February 12-13, and with the CTIO
60$^{\prime\prime}$ telescope on 1997 February 3-5.  We used
long-slit apertures of 2-3$^{\prime\prime} \times$
180$^{\prime\prime}$ and gratings with 300 gpm. The spectral
resolutions are 6\AA\ and 9\AA\ for the MMT and
60$^{\prime\prime}$ telescopes, respectively. Wavelength coverage
is about 3500-8000\AA. We took bias, dome flat and twilight sky
frames each night, and the corresponding corrections (bias
subtraction, flat fielding, and illumination correction) were
applied separately to each night of data. At least 2 standard
stars were observed each night for spectrophotometric
calibration. The observing conditions were not photometric,
except for the CTIO run, so the absolute calibration is subject
to a significant uncertainty. However, relative intensities (such
as a line intensity ratio and an optical power-law index) are
accurate within 20\%, as confirmed by multiple observations of
the same source. Six sources are of undetermined nature because
they are too faint and so gave spectra of too poor a
signal-to-noise.

\centerline{\em 3.2 Classification of Spectra}
\medskip

Of the 45 sources observed at good signal-to-noise, we have
identified 7 red quasars (Table 1).  The results for these 7 red
quasars are presented in this paper. (The full data set will
appear elsewhere.)  The red quasars are mixed in with 18 stars
and a small number of normal blue quasars, narrow emission line
galaxies, and elliptical galaxies (Table 1). The elliptical
galaxies are likely to be brightest cluster galaxies. We shall
report on these separately.  For the remaining 9 X-ray sources,
the red optical candidate within the error circle turned out to
be a star (mostly late type), but it is not likely that these red
stars are the counterparts because their $\alpha_{OX}$ values are
too large for a star (see above, and Maccacaro et al. 1988). The
remaining optical candidates are not red, and hence we stopped
making further observations. These 9 sources, and the 3 blue
quasars measure the blue contamination of the sample, and should
not be considered as part of the list of red X-ray counterparts.

\begin{table}
\caption{Summary of Optical Identifications of Red X-ray Source
Counterparts} 
\begin{tabular}{|lccccccccc|}
\hline
&Total&Red    &Too  &NLXG&Elliptical&M    &Other&Blue   &Not\\
&     &Quasars&Faint&    &Galaxies  &Stars&Stars&Quasars&Red\\
\hline
1.8$>\alpha_{OX}>1.6$&22&1&0&0&1&17&1&0&2\\
(O$\geq$19)&&&&&&&&&\\
\hline
$\alpha_{OX}<1.6$&29&6&6&2&4&1&0&3&7\\
(O$<$19)&&&&&&&&&\\
\hline
Total&51&7&6&2&5&18&1&3&9\\
\hline
\end{tabular}
\end{table}

The optical spectra of the 7 red quasars are shown in Figure 2a,
b. Broad lines of $H\alpha$, $H\beta$ and $MgII$ are clearly seen
in the spectra as well as bright narrow lines (e.g.,
[OIII]$\lambda$5007), making the AGN character of the objects
unambiguous.  In Table 2, we tabulate source position, redshift,
optical magnitude and color, X-ray flux and X-ray colors, and
$\alpha_{OX}$ as well as offsets between the optical and X-ray
positions.

In Figure 1b, each class of identified sources are plotted in the
$\alpha_{ox}~-~(O-E)$ plane. The distribution of these sources
can be compared with the EMSS sources in Figure 1a, confirming
that most X-ray sources with low $\alpha_{ox}$ are indeed quasars
or galaxy clusters. The selection technique finds 7/51 red
quasars, i.e. 8\% efficiency. A slightly stricter criterion,
$\alpha_{ox} < 1.6$ (instead of 1.8), would have selected red
quasars more efficiently (Table~2, Figure 1b): only one M star
(instead of 19 stars) would have been present, with only one red
quasar lost, i.e. 6/29, a little over 20\%. Of the initial 87,
71\% (62) remain when $\alpha_{OX}$=1.6 is the boundary.

\begin{table}
\caption{Basic Properties of Red Quasars}
\medskip
\begin{tabular}{|lccrllllcll|}
\hline
name& RA$^a$ & DEC$^a$& offset$^b$& z& $f_X^c$& $\alpha_{soft}^d$& $\alpha_{hard}^e$&
O&O-E&$\alpha_{OX}$ \\
\hline
\multicolumn{11}{c}{($\alpha_{opt}< -2.0$)}\\
J2255.5+0536 & 22 55 31.0& 05 36 01& 11.3& 0.0647 & 7.74& 1.459& 0.750& 16.38& 2.44& 1.636\\
J1234.3+2614 & 12 34 21.8& 26 13 28&  5.0& 0.3120 & 2.42& 0.860& 1.711& 20.55& 2.35& 1.180\\
J1218.1+2956 & 12 18 07.1& 29 55 21&  4.3& 0.1514 & 1.62& 0.959& 1.512& 19.16& 2.08& 1.466\\
J0909.7+4302 & 09 09 43.6& 43 02 47&  7.0& 0.2748 & 1.65& 0.859& 1.417& 21.38& 2.28& 1.114\\
J1143.6+5521 & 11 43 35.5& 55 20 21&  4.0& 0.1467 & 1.52& 0.535& 1.040& 19.33& 2.33& 1.451\\
\hline\multicolumn{11}{c}{$-0.9 > \alpha_{opt} > -2.0$)}\\
J1051.4+3358 & 10 51 28.3& 33 58 04&  8.0& 0.1829 & 2.76& 1.290& 1.295& 18.27& 2.72& 1.515\\
J1142.6+4624 & 11 42 41.2& 46 24 21&  3.1& 0.1151 &15.75& 0.926& 1.263& 16.33& 2.44& 1.522\\
\hline
\end{tabular}

$a.$ RA and DEC: optical coordinate in Equinox J2000. \\
$b.$ offset: difference of X-ray and optical positions in arcsec. \\
$c.$ $f_X$ in units of $10^{-13} erg s^{-1} cm^{-2}$.\\
$d.$ $\alpha_{soft}$ : X-ray spectral index in 0.1 - 0.8 keV. \\ 
$e.$$\alpha_{hard}$ : X-ray spectral index in 0.8 - 2.0 keV. \\
\end{table}

\medskip
\centerline {\em 3.3 Observed Optical Properties of the 
Red Quasars} 
\medskip

We measured the optical continuum slopes by fitting a power law
to the continuum spectra, after excluding the strong emission
lines (Table 3). Although all the sources were selected based on
a red $O-E$ color, in some cases the observed optical continuum
shape is relatively flat. This is both because line emission
contributed to the blue and/or red bands, and because of
uncertainties on the O and E magnitudes, particularly when the
object is faint (M. Irwin, private communication).  The power law
index ($F_{\nu} \sim\nu^{\alpha_{opt}}$) ranges from $-$0.9 to
$-$2.6.  A steep optical continuum, $\alpha_{opt} < -2$, is found
in 5 of the 7 red quasars, while even the remaining 2
`intermediate' red quasars are redder ($-1.5 < \alpha_{opt} <
-1$) than found for UV excess selected quasars ($-0.2\pm 0.8$,
Neugebauer et al., 1987).  To illustrate the spectral differences
between these two groups, we display the spectra separately in
Figure 2a (steep) and Figure 2b (intermediate).

In addition to the difference in continuum shape, these two
groups also differ in their $H\beta$ line strengths (see Figure
2a, b). The group with the steep optical continuum have only weak
$H\beta$ lines or no detection, whereas the group with a
relatively flat continuum have stronger $H\beta$ lines.  Since
the ratio of $H\alpha$ to $H\beta$ is sensitive to optical
extinction, this suggests more reddening in the `steep slope'
group than the `intermediate slope' group (Table~3), in accord
with the optical continuum slopes.

To quantify this effect for those quasars with no detected
$H\beta$ line, we estimated its upper limit using a simple method
that assumes a box profile with a base equal to
3000~km~sec$^{-1}$ (the mean FWHM of detected $H\beta$ lines),
and a height equal to three times the fluctuation noise on the
continuum. This is a conservative measurement because the peak of
a Gaussian profile would be more easily detected than the flat
top of a box profile, particularly when the line width is
considerably larger than the spectral resolution. Monte Carlo
simulations using a Gaussian line profile assuming Poisson
statistics show that the box profile overestimates the upper
limit by up to 50\% for the adopted line width, while it
reproduces consistent results when the line width is comparable
to the spectral resolution.  For the two objects whose optical
spectra do not cover the $H\alpha$ line we have instead used the
[OIII]/$H\beta$ ratio as a measure of relative $H\beta$ strength.

For all 5 quasars with $\alpha_{opt} > 2$, the $H\alpha/H\beta$
ratios are greater than 5, while the [OIII]/$H\beta > 0.8$.  The
line ratios of the two remaining, quasars are
smaller (Table 2), consistent with less reddening in intermediate
slope objects.

None of the characteristic galaxian stellar absorption features
\footnote{CH g-band$\lambda$4304, MgI$\lambda$5175,
Ca+Fe$\lambda$5269, Na$\lambda\lambda$5890,5896.}
are seen in our spectra. Most strikingly no 4000\AA~ break is seen
in any of the five red quasars for which our spectra cover that
region, including all of the steep slope group.  Typical values
of $D(4000)$
\footnote{$D(4000) = F_{\nu}(4050-4250\AA)/F_{\nu}(3750-3950\AA)$,
Dressler \& Shectman(1987).}
are 1--1.2, as expected from the measured optical slopes. These
compare with values of 2$\pm$0.2 for normal E and S0 galaxies
Dressler \& Shectman(1987).  Hence any starlight continuum
contribution to the red quasar continuum must be minor.

\begin{table}
\caption{Line, Continuum Properties and Luminosities of Red
Quasars {\sc add D4000?}} 

\medskip
\begin{tabular}{|llllllll|}
\hline
name& $\alpha_{opt}^a$&\multicolumn{3}{c}{H$\alpha$}& M(O) &M(E) & L(X)$_{43}^b$\\
    &                 & FWHM & f$^c$ & L$^d$ & & & \\
\hline
\multicolumn{7}{c}{($\alpha_{opt}< -2.0$)}\\
J2255.5+0536 & $-$2.39 &5962 &29.8 &0.54 &$-$21.6 &$-$24.1 &1.02  \\
J1234.3+2614 & $-$2.64 &7184$^e$ & 5.1$^e$ &2.70$^e$ &$-$21.1 &$-$23.5 &8.19  \\
J1218.1+2956 & $-$2.36 &5011 & 5.2 &0.56 &$-$20.8 &$-$22.9 &1.21  \\
J0909.7+4302 & $-$2.43 &2178 & 2.3 &0.91 &$-$20.0 &$-$22.3 &4.29  \\
J1143.6+5521 & $-$2.11 &3223 & 5.1 &0.51 &$-$20.5 &$-$22.9 &1.07  \\
\hline
\multicolumn{7}{c}{($-0.9 > \alpha_{opt} > -2.0$)}\\
J1051.4+3358 & $-$0.93 &1797$^f$ & 7.1$^f$ &7.10$^f$ &$-$22.1 &$-$24.8 &3.07  \\
J1142.6+4624 & $-$1.41 &3104 &81.3 &4.90 &$-$23.0 &$-$25.4 &6.75  \\
\hline
\end{tabular}

$a.$ $\alpha_{opt}$  : optical spectral index,$f_{\nu}\propto\nu^{\alpha_{opt}}$. \\ 
$b.$ $L(X)_{43}$ in units of 10$^{43}$erg~s$^{-1}$.\\
$c.$  $f(line)_{15}$ in units of 10$^{-15}$erg~cm$^{-2}$s$^{-1}$.\\
$d.$  $L(line)_{42}$ in units of 10$^{43}$erg~s$^{-1}$.\\
$e.$ MgII.\\
$f.$ H$\beta$.
\end{table}

\bigskip
\centerline {\em 3.4 X-ray Colors of the Red Quasars}
\medskip

To determine the rough X-ray spectral properties of the 7 red
quasars we first double-checked in the PSPC images that the
sources were cleanly separated from any confusing sources, and
then measured their X-ray hardness (HR=H/M) and softness (SR=S/M)
ratios based on the count rates in the standard ROSAT PSPC bands:
Soft, S (0.1-0.4~keV), Medium, M (0.4-0.86~keV), and Hard, H
(0.87-2~keV). These ratios are then converted to effective X-ray
spectral indices $\alpha_{soft}$ and $\alpha_{hard}$ (Table 2),
to correct for the variable Galactic line-of-sight absorption,
and the energy-dependent PSF. [These are not physical slopes, but
should be considered analogous to U-B and B-V colors, see Fiore
et al. (1998) for a detailed discussion of the estimation and
usage of effective X-ray spectral indices.]  Due to the low
signal-to-noise ratio of X-ray data, individual spectral indices
are not reliable.  However, the locus of their colors forms a
useful indicator of global X-ray properties.

We compare the colors of the red quasars with those of normal
radio-quiet quasars in Figure 3. The large filled symbols are the
red quasars reported here, while the cloud of small dots are
radio-quiet quasars from the sample of Fiore et al. (1998). On
average the red quasars have smaller $\alpha_{soft}$ than
$\alpha_{hard}$, indicating a cut-off spectral shape. [The line
pairs around the periphery of the figure show outline spectral
shapes for their locations in the
($\alpha_{soft}$,$\alpha_{hard}$) plane.] This is consistent with
their having the moderate X-ray absorption expected from their
optical properties (Fiore et al., 1998a).

\bigskip
\centerline {\bf 4. Discussion}
\medskip

These observations show that a population of red AGN can be
extracted efficiently from the ROSAT pointed archive. Moreover
the red AGN we find are radio-quiet. None of them is a radio
source in the VLA NVSS survey ($f_{1.4GHz}<$2.5~mJy, Condon et
al., 1998) implying $R_L=log[f(opt)/f(5GHz)]<2.0$, compared with
$2< R_L<5 $ for radio-loud quasars (Wilkes \& Elvis 1987).  The
agreement of X-ray colors, optical continuum slope and
H$\alpha$/H$\beta$ ratios with the same value of obscuring dust
and gas (A$_V\sim$1-2) argues for their being dust reddened
objects rather than instrinsically red continua.

Puchnarewicz \& Mason (1988) discuss a similar population of 14
candidate red, $\alpha_{opt}>2$, AGN derived from the `RIXOS'
sample, which extends to several times fainter X-ray fluxes. The
RIXOS sample was selected from the ROSAT pointed archive based on
X-ray flux alone. (There is one object in common between the two
samples.)  Two of the red RIXOS AGN may be intrinsically red, and
three have clear reddening. So there is currently a total of 8
reddened AGN available from ROSAT.

The red AGN, both from this sample and from RIXOS, are borderline
quasar/Seyfert objects.  The observed optical luminosities of the
red AGN are modest, lying at the high end of the traditional
Seyfert luminosity range ($M_B> -23$ mag, Veron-Cetty \& Veron
1984, Schmidt \& Green 1983): from $M_O=-$20 to $M_O=-$22 mag
\footnote{We have used $H_0=50$ km sec$^{-1}$~Mpc$^{-1}$ and 
$q_0= 0$.}.
The dereddened, intrinsic luminosity of the sources is likely to
be significantly higher, depending on the amount of absorption
(the lower limit on the extinction is 2-3 mag in O). This places
them at $-25<M_O<-22$, within the quasar regime. Similarly, the
observed X-ray luminosity ranges from $1.3 \times 10^{43}$ to
$1.2\times 10^{44}$ erg sec$^{-1}$, while the intrinsic
luminosities are likely to be higher by about a factor of 2,
depending on the intrinsic spectrum and the amount of absorption
present.

The precise allocation of these AGN as quasars or Seyferts is not
fundamentally very interesting, since this is just a conveniently
chosen value on a continuous luminosity scale. For simplicity we
will refer to them as `red quasars' for the rest of this paper.
It is, however, interesting to understand why much more, or much
less, luminous AGN were not found in the two red quasar searches.
Is this a selection effect, or is there a physical preference for
red objects to cluster in a limited range of luminosity? We shall
return to this question later (\S 4.3).

The existence of a red quasar population immediately raises
important questions: What makes them red, compared to usual blue
quasars? How common are they, particularly once allowance is made
for their reduced flux due to probable obscuration?  If they are
absorbed, then by how much?  Where is the absorbing dusty
material? Might a larger, more obscured, AGN population exist?

\medskip
\centerline{\em 4.1 How Common are Red Quasars?}

We can address the relative numbers of red quasars in a rough
way.  The population, although a minority, is quite common. We
found 7 red quasars out of 45 candidates for which our
spectroscopy was adequate to produce a classification.  Assuming
the 45 were a random subsample of the original sample of 87, then
that sample would produce 14 red quasars.  This is 2.4\% of 575
sources with optical colors available.  At our flux limit Stocke
et al. (1991) find that 51\% of all X-ray sources are AGN.  So a
minimum of 13.5/288 = 4.7\% of AGN in our `unidentified sample'
are radio-quiet red quasars (with O$<$20) at this soft X-ray flux
level.  The Boyle \& di~Matteo (1995) upper limit of 9\% of CRSS
X-ray sources being red quasars is consistent with the minority
population of moderately obscured (A$_V$=2) quasars we have found
here.

Additional red quasars could be hidden in our sample. Our sample
of 45 classified spectra is not a random subsample of the
candidate list. Figure 1b shows that we preferentially selected
objects with $\alpha_{OX}>1.3$, i.e. the brighter objects (with
B$<$21.5).  If the 6 objects for which we attempted to get
spectra turn out to be red quasars then the fraction of red
quasars among the PSPC AGN could be almost double our first
estimate. There are a further 36 red candidates for which no
spectroscopy was attempted.  So the true occurence rate of red
quasars is uncertain by a factor 7.

To find the fraction of red quasars among all the AGN in our
X-ray flux limited sample we must allow for the 165--300 blue AGN
in the `identified' sample, so 1\% -- 7\% of the whole soft X-ray
AGN population is red.

If the red AGN are obscured then the intrinsic rate of occurence
of red quasars has to be calculated relative to their unreddened
parent population.  Obscuration by $A_V=2$ reduces their
unobscured X-ray fluxes by a factor $\sim$2.  Since higher flux
AGN are rarer (the X-ray selected AGN logN-logS relation has a
slope of $-$3/2 in this flux range, Hasinger 1992, Della~Cecca et
al., 1992), the red population forms a larger fraction of this
population, $\sim$3\% -- 20\%. 

Using the fraction of the sources at the intrinsic flux of the
red quasars allows us to compare our result with samples
unaffected by obscuration, and with model predictions.
Comparisons with the observed frequency of radio-loud red quasars
with broad emission lines found by Smith and Spinrad (1980,
178~MHz 3CR) and Stickel et al. (1996, 5GHz ``1Jy'' sample), and
with the Comastri et al. (1995) predictions are straightforward.

Red quasars are found in 15\% of the 3CR source sample and
6\%-20\% of the 1~Jy (Carilli et al., 1998) sample. These are
comparable with the 3\% - 20\% of the `intrinsic flux'
X-ray population that we find, suggesting that the radio-loud and
radio-quiet quasar populations have similarly sized populations
of moderately obscured quasars.
The obscuration in the 3CR red quasars is also about A$_V$=2
(Elvis et al., 1994, Economou et al., 1995, Rawlings et al.,
1995). For four objects in the 1Jy sample Carilli et al. (1998)
estimate lower limits on A$_V$ from 2 to 5, based on
extrapolating the radio-infrared index to the optical. However,
since such steeply rising slopes are not known among unobscured
quasars, these limits are likely to be too large, and values
comparable with the 3CR estimates are probably acceptable.

For AGN with $N_H<10^{22}$cm$^{-2}$ the Comastri et al.(1995)
model predicts that 26\% will have $10^{21}<N_H<10^{22}$cm$^{-2}$
($A_V\sim 2$), somewhat larger but comparable with the numbers
found here. Comastri et al. predict far larger numbers of more
obscured objects.

\medskip
\centerline{\em 4.2 More Obscured Objects}

More obscured objects may exist. Puchnarewicz \& Mason (1998)
find several objects with steeper optical continua, and figure~1b
shows several redder candidates and many more x-ray loud
candidates, with no optical spectra to date.  Webster et
al. (1995) suggested that A$_V$=5 may be typical of their red
objects, giving their putative ROSAT counterparts in our sample
V=23-25 which is below the Palomar Sky Survey limit. A$_V$=5
corresponds to a column of 9$\times$10$^{21}$atoms~cm$^{-2}$,
which would reduce ROSAT PSPC count rates to 15\% of their
unobscured values. These objects would then be hidden as a 6\%
minority among the more common lower luminosity unabsorbed
quasars if they had the same {\em unobscured} space density as
normal blue quasars.  Some 9\% of our initial X-ray selected
sample are `blank field' objects, i.e. have no counterpart on the
Palomar Sky Surveys. A fraction of these could be more heavily
obscured red quasars.  Boyle \& di Matteo (1995) find that the
CRSS sample could be missing no more than 9\% (at the observed
flux) in red quasars.  Subtracting the 1\% of moderately
obscured quasars that we have identified this still leaves 8\%
that could be highly obscured. Hence the CRSS result is, perhaps
surprisingly, consistent with a sizeable, heavily obscured,
population. The Comastri et al. (1995) model predicts a
comparable population of AGN with N$_H\sim$10$^{22}$cm$^{-2}$,
1.1 times larger than the unobscured population.

In our sample the occurrence of red quasars appears to be four
times higher for O~$>$~19 mag (4/12) than for O~$<$~19 mag 3/32)
(Table~2), although the number of sources is small. Such a trend
is expected if the quasars are heavily absorbed.  An increase in
N$_H$ from 3$\times$10$^{21}$cm$^{-2}$ to
1$\times$10$^{22}$cm$^{-2}$ cuts the ROSAT flux by a factor 2.4,
but reddens the V-band by 3.8~magnitudes (a factor 33). So the
optically fainter sources might well be redder. However the RIXOS
red quasars (Puchnarewicz \& Mason 1998) show no correlation of
optical slope with $m_V$ - the three steepest slopes are all in
the brighter half of the sample of 14.  Gunn \& Shanks (1998)
have pointed out that, while redshifting the ultraviolet into the
optical increases the effects of reddening, the corresponding
shift of hard X-rays into the soft ROSAT band {\em decreases} the
effectiveness of reddening.

To estimate an accurate fraction of this potential hidden
population needs a larger sample, including more absorbed,
fainter objects.  At even larger column densities
(10$^{23}$-10$^{24}$cm$^{-2}$) the Comastri et al.(1995) model
predicts nearly four times the unobscured population.  More
heavily obscured quasars could be found in hard X-ray surveys,
from ASCA (Ueda et al., 1998), and the Beppo-SAX `HELLAS' survey
(Fiore et al., 1999).

\medskip
\centerline{\em 4.3 Limited Luminosity Range}

It is striking that while ROSAT surveys that are defined simply
by an X-ray flux limit find AGN spanning over 3 decades in X-ray
luminosity (e.g. `CRSS' Boyle, Wilkes \& Elvis 1997, RIXOS,
Puchnarewicz et al., 1996), both RIXOS and this survey find red
quasars in only one decade of luminosity, and that this decade is
the lowest one in which RIXOS and CRSS AGN are found.  A 2-tail
Kolmogorov-Smirnov test shows that the chance that red and
non-red AGN from RIXOS come from the same luminosity distribution
is only $\sim$2\%.  This suggests that predominantly lower
luminosity AGN are obscured. [Note that the observed amount of
obscuration only decreases the observed X-ray luminosity by a
factor of $\sim$2, and so does not itself cause the low observed
luminosities.]

Similar suggestions have been made before: Lawrence \& Elvis
(1982) found that only AGN below L$_X\sim 10^{44}$erg~s$^{-1}$
(2-10~keV) showed obscuration. Occasional examples of highly
obscured `type 2' (i.e. narrow line) quasars have been reported
(Stocke et al., 1982, Almaini et al., 1995, Shanks et al., 1995),
and careful searches have found broad H$\alpha$ in most cases
(Halpern, Eracleous \& Forster 1998), making them similar to the
red quasars found here. Searches among the fainter objects in our
sample and searches at higher energies (e.g. the BeppoSAX
`HELLAS' survey, Fiore et al., 1999) will be more effective at
finding a high luminosity red quasar population.

The Comastri et al. X-ray background models assume luminosity
functions for the obscured objects that are identical to those of
the unobscured objects, except for normalization and so predicts
high luminosity red quasars. If instead obscured AGN occur
preferentially at low luminosity this will substantially affect
the model predictions. We would, for example, expect the obscured
population to be more numerous and at lower redshift.

If there is a real deficit of high luminosity red quasars, then
one possibility to explain this lack might have been that as an
AGN became more luminous the continuum ionized the obscuring
medium, rendering it transparent to X-rays. However ionized
absorbers are also more common at lower luminosities (Laor et
al., 1994). So most likely high luminosity AGN have fewer lines
of sight with intervening material regardless of ionization
state. (Interestingly this is in the same sense as the Baldwin
effect: that higher luminosity quasars have weaker CIII]1909
emission lines.)  Any physical model of a quasar would need to
explain this difference.

\medskip
\centerline{\em 4.4 Physical Properties}

We can say only a little about the physical properties of the red
quasars from this data. 

The consistency of the optical reddening indicators with the
X-ray colors suggests that the same obscuring material covers
both emitting regions, and that it lies outside the broad
emission line region.

Smith and Spinrad (1980) suggested that the redness of the red
3CR quasars is intrinsic to the continuum emission process, based
on the lack of an absorption feature at $\lambda$ =2200\AA, which
is a typical characteristic of Milky Way dust (eg., Bless and
Savage 1972). The detection of 21~cm HI absorption toward a large
fraction of the 1Jy (Stickel et al., 1996) red quasars (Carilli
et al., 1998) argues for dust reddening in those objects. Since
our sample of radio-quiet quasars is relatively nearby (with
redshifts up to 0.3), we can not check for this feature directly.
However, this explanation is in contrast to the observed Balmer
decrement, and the X-ray colors.  It is possible that the
reddened 3CR sources contain dusty, ionized absorbers, as seen in
3C212 (Mathur 1994; Elvis et al. 1994) and IRAS17020+4544
(Komossa \& Bade 1998), where the dust composition may differ
depending on, for example, the quasar continuum
shape. Ultraviolet observations are needed to investigate the
$\lambda$ =2200\AA\ feature, but are probably infeasible at
present.

\medskip
\centerline{\em 4.5 Other Red AGN}

Some previous studies have considered red AGN-like objects in
X-ray surveys: 

The RIXOS survey (Puchnarewicz \& Mason, 1998) found that 9\%
(14/160) of their AGN were red. However, only 3 of the RIXOS
sources have Balmer decrements that require reddening, so the
true occurence rate of reddened objects may be similar to that
which we have found.  The fainter flux limit of the RIXOS survey
may render more absorbed objects visible. Certainly, the steeper
optical slopes ($2.5<\alpha_{opt}<4.0$) of half the RIXOS sample
suggest greater reddening.

Kruper and Canizares (1989) studied red AGNs in {\em Einstein}
X-ray selected samples and indirectly concluded that these are
red due to the presence of host galaxies. Benn et al. (1998)
arrive at a similar conclusion for low frequency selected
radio-loud quasars.  However no galaxian starlight features are
seen in our spectra and the host galaxy cannot explain the
observed Balmer decrements in our sample.  The Kruper \&
Canizares objects are not as red as our samples, having B-I= 1.5
to 2.5 mag. If we take R-I to be 0.5 - 1.0 mag (this is the R-I
range of the samples in their table 2), B-R would be less than
2.0, our defining threshold. In fact none of their objects with
measured R magnitudes exceed B-R = 2.0. 

The Narrow Line X-ray Galaxies (NLXGs) found plentifully in deep
ROSAT surveys (e.g. Boyle et al., 1995a), are also normally
assumed to be obscured AGNs (e.g. Hasinger et al. 1998, Schachter
et al., 1998).  X-ray absorbed NLXGs could contribute
significantly to the cosmic X-ray background if they are more
common at fainter X-ray flux levels, as suggested by McHardy et
al. (1998, see also Hasinger et al. 1998 for a cautionary note).
Although they have similar X-ray luminosities to the NLXGs, the
X-ray selected red quasars are not simply the same population,
however.  The red quasars have the normal broad optical emission
lines of quasars, while NLXGs have either none, or only extremely
weak ones (Boyle et al., 1995b, c.f. our Figure 2a,b).  Moreover,
NLXGs usually exhibit blue optical continua (for example, O-E $<$
2 for 5 out of 6 NLXGs in CRSS, Boyle et al. 1997), while red
quasars have red optical continua (O-E $>$ 2). Further X-ray and
optical study of these objects may let us understand whether they
are two separate populations or are related by e.g. special
viewing geometry or scattering of a blue continuum.


The Palomar survey of the nuclei of bright galaxies (Ho,
Filippenko \& Sargent, 1997) is based on a sample of galaxies
selected for their non-nuclear properties and so is less biased
against finding red AGN than most other optical search methods,
The Palomar AGN are low luminosity AGN, allowing us to see if the
high incidence of red AGN at lower luminosities continues to
increase at even lower values.  The Palomar survey finds that
18\% (8/44) of Seyfert nuclei have A$_V>$2, based on their narrow
line Balmer decrements (Ho, Filippenko \& Sargent 1997). However
almost all of these are LINERs or type~2 Seyferts, which may have
large reddening toward the broad line region.. Only one Seyfert
(NGC7479) shows any evidence for a broad line component. Broad
emission lines are extremely hard to detect at these flux levels,
but the suggestion is that the middling luminosities, toward the
Seyfert/quasar borderline are particularly prone to moderate
obscuration.

\bigskip
\centerline{\bf 5. Conclusions}

Radio-quiet red quasars can be found in substantial numbers. They
comprise at least 1\%, and potentially 7\%, of the soft X-ray
population in a flux limited survey. Correcting the X-ray fluxes
to their intrinsic values puts them among brighter AGN where they
form 3\% of the population. Allowing for `blank field' sources
as much as 20\% of ROSAT selected quasars may be red, at a given
unobscured flux. The size of this population is consistent with
previous upper limits, with the Comastri et al. (1995) model for
the X-ray background, and with the size of the radio-loud 3CR and
`1Jy' red quasar populations. Red quasars seem to be
preferentially lower luminosity objects on the Seyfert/quasar
borderline, but not at higher or lower luminosities. Such a bias
against obscured high luminosity objects would affect X-ray
background estimates for this population, and would need
explaining in a physical model of quasars.

We stress that the quasars we find have broad optical lines. They
are not Narrow Line X-ray Galaxies which, by contrast, have
predominantly narrow optical permitted lines and blue continua.

The optical slopes, H$\alpha$/H$\beta$ ratios, and X-ray colors
are all consistent with reddening by A$_V\sim$2, assuming
standard Milky Way dust properties. So the same obscuring
material probably covers each of the emitting regions.

A significant population of more highly obscured (A$_V$=5)
quasars could well exist and be consistent with the results here,
with earlier ROSAT limits, and would be as predicted by the
Comastri et al. (1995) model. Hard X-ray surveys will soon settle
the question of the size of any such population.

Using a minor refinement of the technique presented here, red
quasars can be found with high (20\%) efficiency in the ROSAT
data. 

\bigskip
\bigskip
\centerline {Acknowledgment}

We thank L. Angelini for helping us to use WGACAT and M. Irwin
for user of the APM on-line catalog. We also thank F. Fiore and
F. Nicastro for providing their program to calculate X-ray
spectral indices, and F. Fiore once again for supplying the X-ray
slopes of the radio-quiet sample in figure 3. The support by the
FLWO, MMT and CTIO staffs in setting up and operating various
instruments were invaluable to this study. The HEASARC/GSFC PIMMS
program was used to calculate PSPC count rates. This work was
supported by NASA grants NAG5-3066 (ADP), NAG5-6078 (LTSA), and
NASA contract NAS8-39073 (ASC).

\newpage
\centerline {Figure Captions}

Figure 1. (a) Optically identified EMSS sources in the
$\alpha_{ox}$ -- (O-E) plane. Different sources are marked by
different symbols.  Our source selection criteria ($O-E > 2$ and
$\alpha_{ox} < 1.8$) are seen as a lower-right box. (b) same as
(a) but for our ROSAT samples. The $\alpha_{OX}$ line shows how a
slightly stricter criterion yields a larger fraction of red
quasars. 

Figure 2. (a) Observed spectra of the 5 radio-quiet red quasars
(O-E $>$ 2 mag and $\alpha_{opt} > 2.0$). Broad lines such as
$H\alpha$, $H\beta$ and $MgII$ are clearly seen in the spectra as
well as bright narrow lines (e.g., [OIII]$\lambda$5007). Those
strong lines are marked in the figures. Steep continuum slopes
and very weak $H\beta$ line strengths indicate a significant
amount of dust extinction ($A_V > 2$ mag), in contrast to (b) two
quasar spectra with $\alpha_{opt} < 2.0$ and relatively strong
$H\beta$ lines.

Figure 3. Effective soft and hard ROSAT PSPC X-ray spectral
indices (see text) of the red quasars (large filled symbols),
compared with normal radio-quiet quasars (small dots, Fiore et
al., 1998). Steep optical spectrum $\alpha < -2$) red quasars are
shown as circles, intermediate optical spectrum $-1.5< \alpha <
-1$) red quasars are shown as triangles.  The line pairs around
the periphery of the figure show outline spectral shapes for
their locations in the ($\alpha_{hard}$,$\alpha_{hard}$) plane.

\begin{figure}[b]
\psfig{file=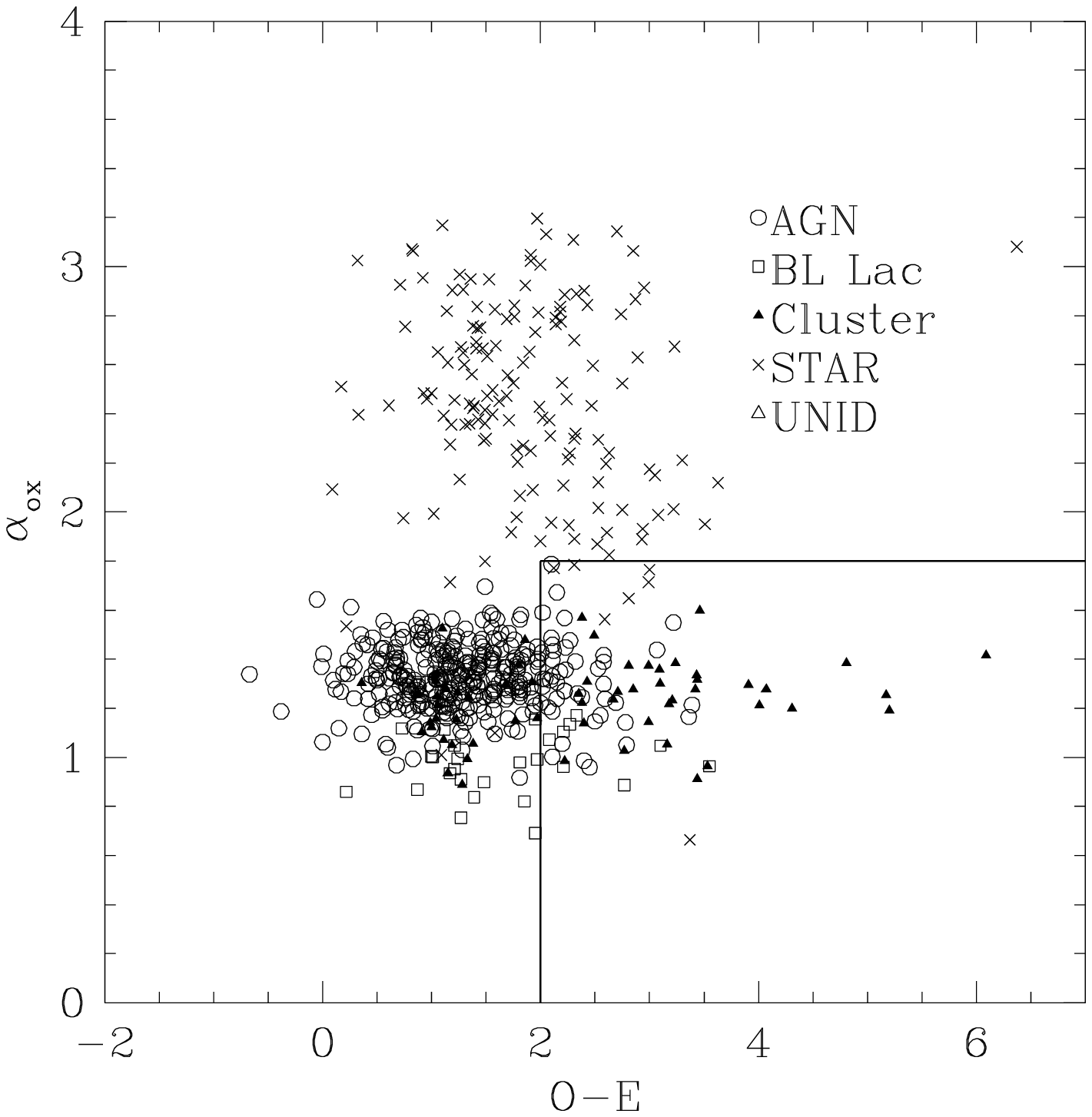,width=6.0in}
\end{figure}

\begin{figure}[b]
\psfig{file=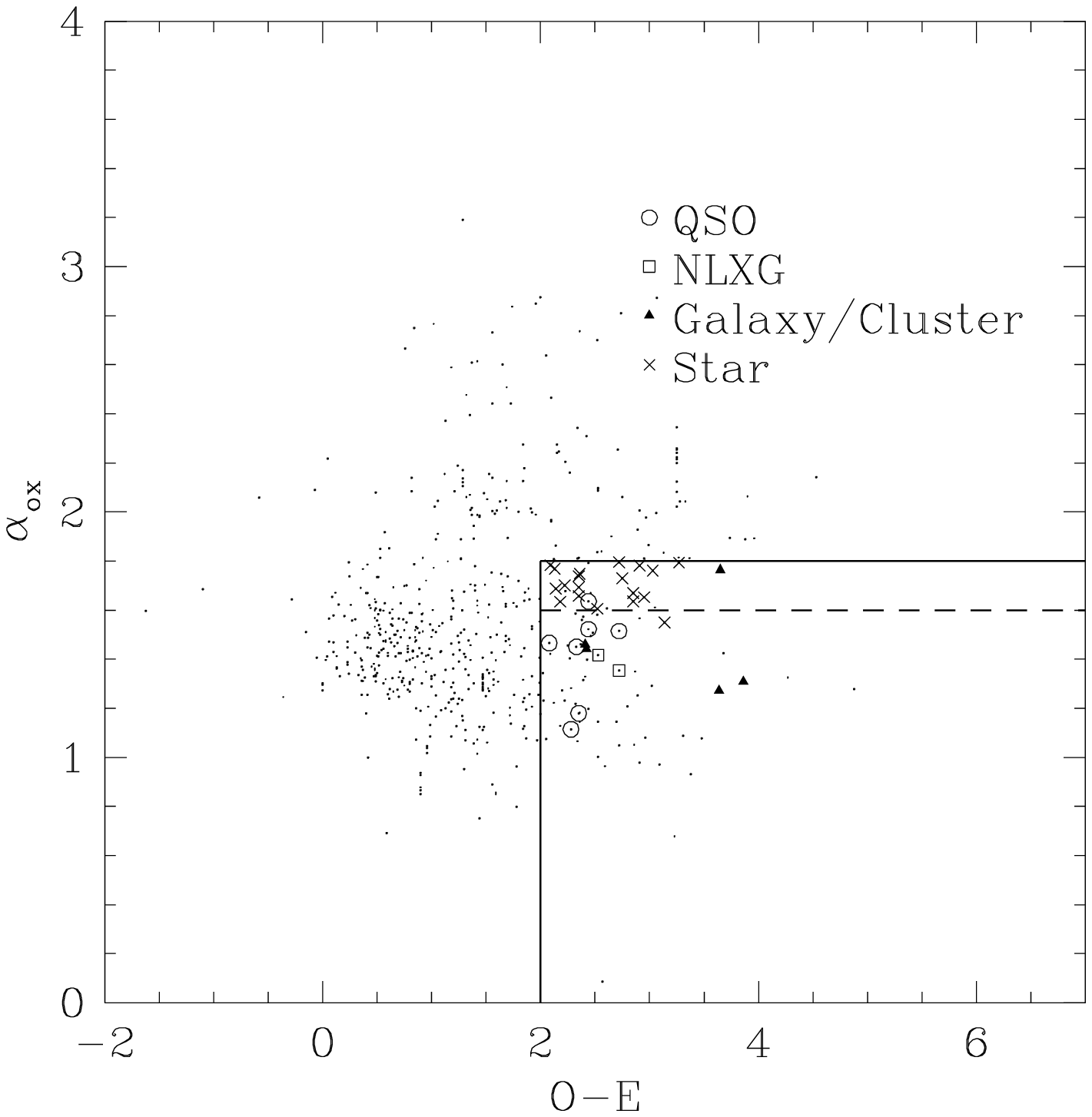,width=6.0in}
\end{figure}

\begin{figure}[b]
\psfig{file=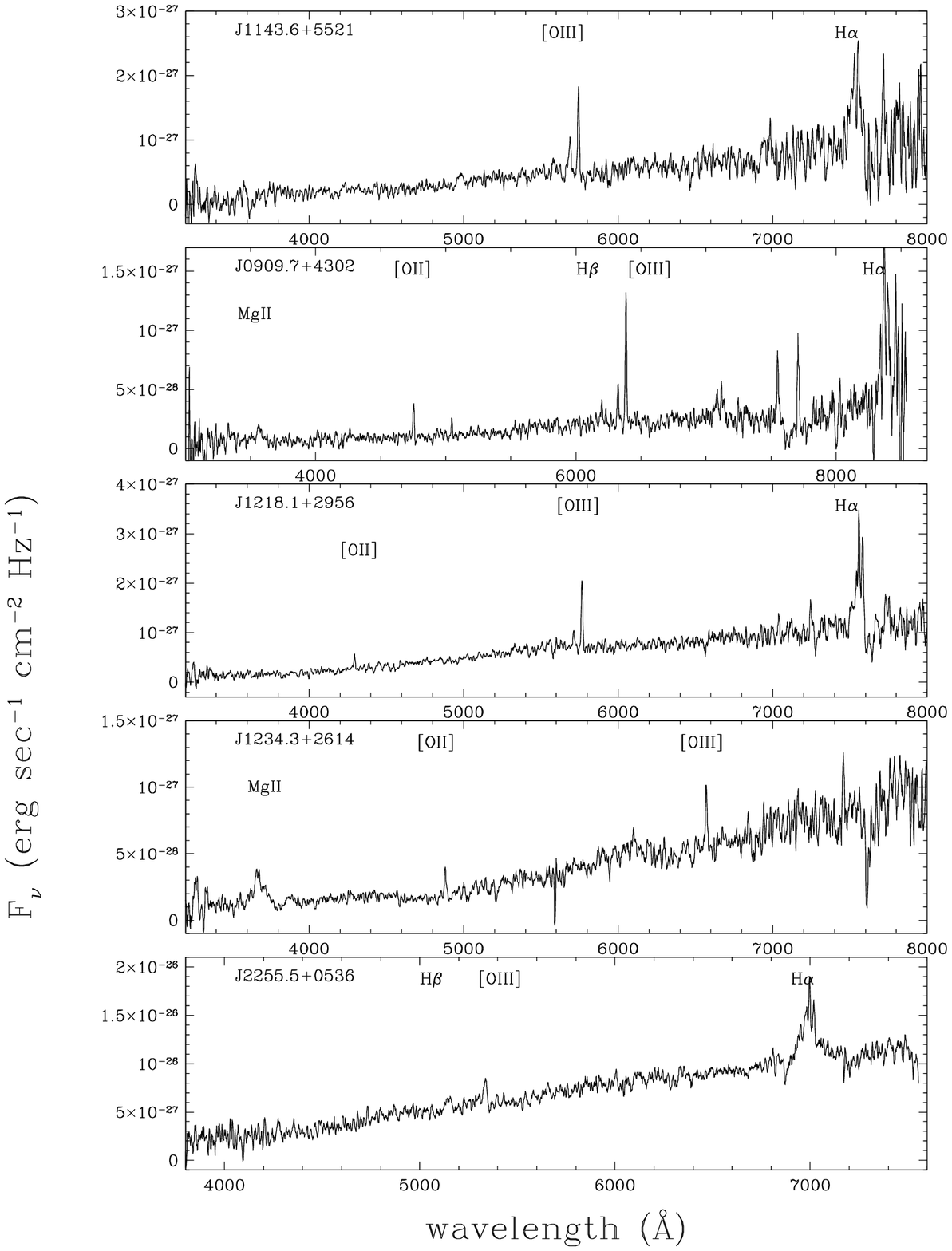,width=6.0in}
\end{figure}

\begin{figure}[b]
\psfig{file=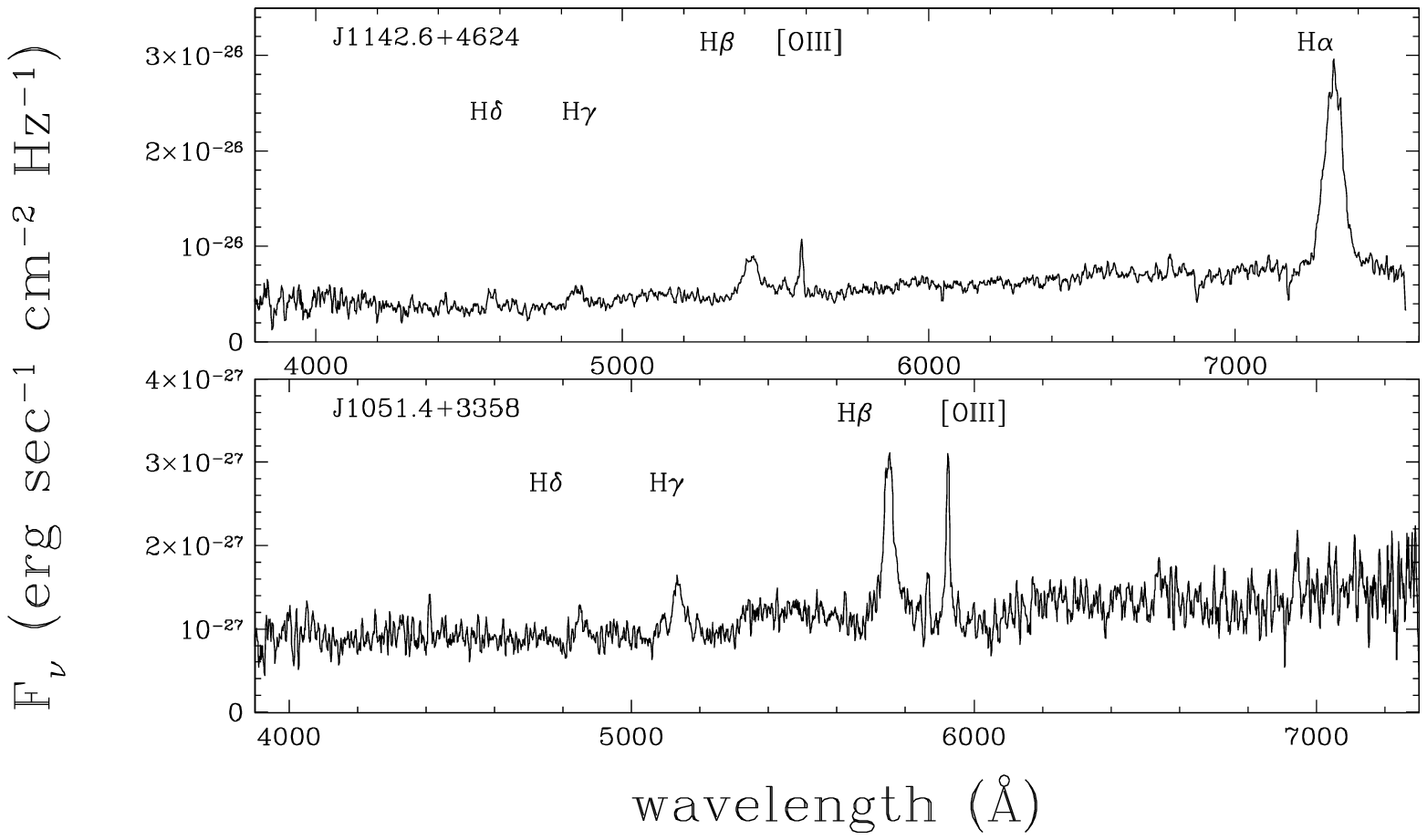,width=6.0in}
\end{figure}

\begin{figure}[b]
\psfig{file=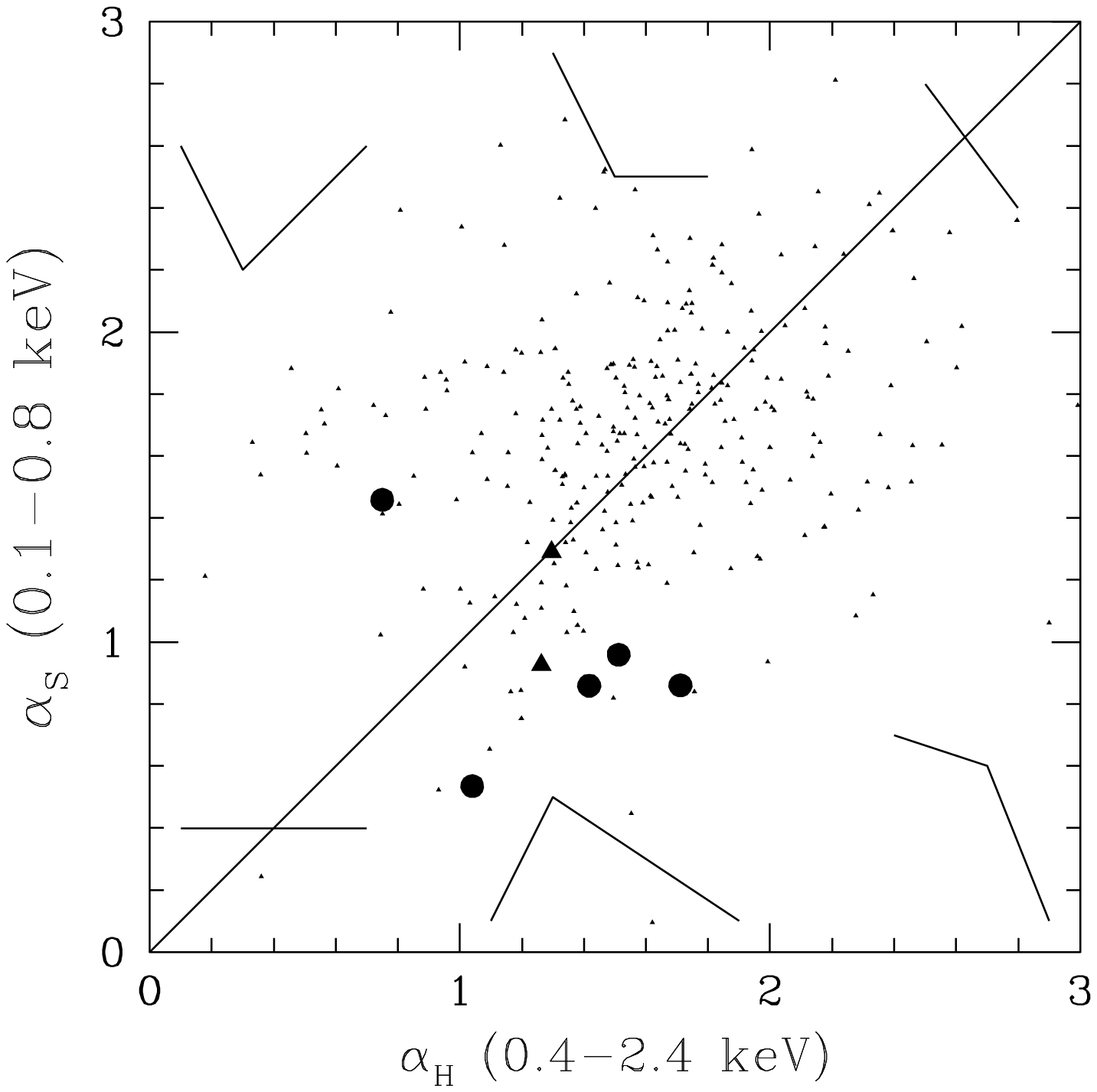,width=6.0in}
\end{figure}

\end{document}